\documentclass[12pt]{article}
\pdfoutput=1

\newlength{\abstractwidth}
\usepackage{amsmath, nccmath}
\usepackage{amsfonts}
\usepackage{amssymb}
\usepackage{latexsym}

\usepackage{color}
\usepackage{graphicx}
\usepackage{tikz}
\usepackage{dsfont}

\usetikzlibrary{arrows,shapes,positioning}
\usetikzlibrary{decorations.markings}
\usepackage[rightcaption]{sidecap}
\tikzstyle arrowstyle=[scale=1]
\tikzstyle directed=[postaction={decorate,decoration={markings,
    mark=at position .65 with {\arrow[arrowstyle]{stealth}}}}]
\tikzstyle reverse directed=[postaction={decorate,decoration={markings,
    mark=at position .65 with {\arrowreversed[arrowstyle]{stealth};}}}]
\usetikzlibrary{positioning}

\usepackage{hyperref}
\definecolor{darkred}{rgb}{0.8,0.1,0.1}
\hypersetup{colorlinks=true, linkcolor=darkred, citecolor=blue, linktoc=page}

\usepackage{cite}

\renewcommand{\thefootnote}{\fnsymbol{footnote}}
\renewcommand{\thanks}[1]{\footnote{#1}}
\newcommand{\starttext}{
\setcounter{footnote}{0}
\renewcommand{\thefootnote}{\arabic{footnote}}}

\newcommand{\bea}{\begin{eqnarray}}
\newcommand{\eea}{\end{eqnarray}}
\newcommand{\be}{\begin{eqnarray}}
\newcommand{\ee}{\end{eqnarray}}
\newcommand{\bma}{\begin{matrix}}
\newcommand{\ema}{\cr\end{matrix}}

\newtheorem{thm}{Theorem}[section]

\def\cC{{\cal C}}

\def\cF{{\cal F}}
\def\cG{{\cal G}}
\def\cH{{\cal H}}

\def\cL{{\cal L}}
\def\cM{{\cal M}}

\def\cO{{\cal O}}

\def\cS{{\cal S}}

\def\ZZ{{\mathbb Z}}
\def\RR{{\mathbb R}}

\def\CC{{\mathbb C}}

\def\QQ{{\mathbb Q}}

\def\Re{{\rm Re \,}}
\def\Im{{\rm Im \,}}

\def\half{{1\over 2}}
\def\thalf{{\tfrac{1}{2}}}
\def\p{\partial}

\def\ep{\varepsilon}

\def\no{\nonumber}
\def\sm{\smallskip}

\begin{document}
\starttext
\setcounter{footnote}{0}

\begin{flushright}
2019 May 15 \\
revised 2019 June 12
\end{flushright}

\vskip 0.3in

\begin{center}

{\Large \bf Integral of two-loop modular graph functions \footnote{Research  supported in part by the National Science Foundation under research grant PHY-16-19926.
}}

\vskip 0.1in


\vskip 0.3in

{\large  Eric D'Hoker} 

\vskip 0.1in

 { \sl Mani L. Bhaumik Institute for Theoretical Physics}\\
{\sl Department of Physics and Astronomy }\\
{\sl University of California, Los Angeles, CA 90095, USA} 

\vskip 0.05in

{\tt \small dhoker@physics.ucla.edu}

\end{center}

\begin{abstract}

The integral of an arbitrary two-loop modular graph function over the fundamental domain for $SL(2,\ZZ)$ in the upper half plane is evaluated using recent results on the Poincar\'e series for these functions.

\end{abstract}

\newpage

\baselineskip=15pt
\setcounter{equation}{0}
\setcounter{footnote}{0}

\newpage
\setcounter{tocdepth}{2}
\newpage

\section{Introduction}
\setcounter{equation}{0}
\label{sec:1}

A modular graph function is an $SL(2,\ZZ)$-invariant function on the Poincar\'e upper half plane~$\cH$ which is associated with a certain type of graph \cite{DHoker:2015gmr,DHoker:2015wxz}.   Modular graph functions provide a natural generalization to higher loop graphs of real-analytic Eisenstein series, which are associated with one-loop graphs.  They are key building blocks for the integrand on the moduli space of genus-one Riemann surfaces of the low energy expansion of superstring theory to genus-one order \cite{Green:1999pv, Green:2008uj}. 

\sm

Modular graph functions obey a wealth of differential and algebraic identities \cite{DHoker:2015gmr,DHoker:2015sve,Basu:2015ayg,DHoker:2016mwo,Basu:2016kli,DHoker:2016quv,Kleinschmidt:2017ege}; their Poincar\'e series was obtained for low weight in \cite{DHoker:2015gmr,Ahlen:2018wng,Dorigoni:2019yoq} and for general weight in \cite{DHoker:2019txf}; the Laurent polynomial part of their constant Fourier mode was derived for low weight in \cite{DHoker:2015gmr,Zerbini} and for arbitrary weight at two-loop order in \cite{DHoker:2017zhq};  the full Fourier series at two-loop order was calculated in \cite{DHoker:2019txf}. The significance of multiple zeta-values \cite{Blumlein:2009cf} in string theory and modular graph functions was explored in \cite{SCHLotterer:2012ny,Broedel:2013tta,Broedel:2014vla,Zerbini,Broedel:2018izr}, while relations with period integrals, mixed motives, and equivariant iterated Eisenstein integrals were studied in \cite{Brown1,Brown2, Broedel:2018izr}.

\sm

String perturbation theory requires, however, not just the integrand but the actual integral of  modular graph functions on the moduli space of genus-one Riemann surfaces $\cM= \cH/SL(2,\ZZ)$. The purpose of this paper is to define and evaluate such integrals for all modular graph functions associated with two-loop graphs. 

\sm

For modular functions which tend to zero at the cusp (cuspidal functions), such integrals may be computed using the Rankin-Selberg method, in terms of the constant Fourier mode of the cuspidal function \cite{Rankin, Selberg}. Modular graph functions, however, have polynomial growth at the cusp, just as real-analytic Eisenstein series do, and their integration requires regularization. A standard regularization is obtained by removing from the domain of integration $\cM$ a small neighborhood of the cusp, and integrating over $\cM_L$ defined for $L >1$ by,
\bea
\cM_L= \cM \cap \left \{ \Im(\tau) < L \right \} 
\hskip 1in 
\cM= \left \{ \tau \in \cH,  ~ |\tau|\geq 1, ~ |\Re(\tau)|\leq \thalf \right \}
\eea 
A generalization of the Rankin-Selberg method, which is applicable to modular functions with polynomial growth at the cusp, was introduced in \cite{Zagier} and evaluates integrals over $\cM_L$ of expressions linear, bilinear, and trilinear in Eisenstein series.  

\sm

The integrals over moduli space required for superstring amplitudes are unique and free of divergences, but their construction requires subtle analytic continuation in the kinematic variables of the amplitude. The existence of this analytic continuation for the four-graviton amplitude at genus one was proven in \cite{DHoker:1994gnm}, where the results of the analytic continuation were used to obtain the mass-shifts and decay widths of massive string states. The evaluation of the amplitude to low orders in the kinematic parameters was carried out in \cite{Green:1999pv, Green:2008uj,DHoker:2015gmr}. The integrals of modular graph functions regularized by integrating over the truncated fundamental domain $\cM_L$ provide a key part of the evaluation of the analytically continued amplitudes.  Recent investigations of the transcendentality properties of the genus-one amplitude  in \cite{DG} directly motivate the problems addressed in the present paper.  

\sm

The organization of the remainder of this paper is as follows. In section \ref{sec:2} we shall briefly review the definition and basic properties of modular graph functions in terms of their Kronecker-Eisenstein and Poincar\'e series representations. The main results of the paper are presented in Theorems \ref{thm1} and \ref{thm2} of section \ref{sec:3}, respectively giving the (regularized)  integrals for two-loop modular graphs functions of arbitrary odd and even weight. We conclude in section \ref{sec:4} with a discussion of open problems and speculations regarding the integrals of higher loop modular graph functions.

\vskip 0.2in 

\noindent
{\bf \large Acknowledgments}

\medskip

We are happy to thank Bill Duke, Michael Green, Justin Kaidi, and Pierre Vanhove for various  collaborations which led to the present paper, and for subsequent discussions.  Also, we are happy to acknowledge the Niels Bohr International Academy in Copenhagen for the warm hospitality extended to the author during part of this work.  This research is  supported in part by the National Science Foundation under research grant PHY-16-19926.

\section{Modular graph functions}
\setcounter{equation}{0}
\label{sec:2}

In this section we shall give a brief review of the definition and basic properties of modular graph functions needed in the sequel of the paper.

\sm
 
A decorated connected graph $(\Gamma, A,B)$ with $V$ vertices and $R$ edges is specified by the connectivity matrix $\Gamma$ of the graph and the decoration $(A,B)$ of the edges of the graph. The components of the connectivity matrix $\Gamma$ are denoted by $\Gamma _{v \, r}$ where $v=1,\cdots, V$ labels the vertices and $r=1,\cdots, R$ labels the edges. No edge is allowed to begin and end on the same vertex: when edge $r$ contains vertex $v$ we have $\Gamma _{v \, r}=\pm 1 $  while otherwise $\Gamma _{v\, r} = 0$. The  decoration $(A,B)$ of the edges is specified by two arrays of {\sl exponents} $a_r,b_r$,
\bea
\label{ab}
A = [a_1, \cdots , a_R]  \hskip 1in  B = [b_1, \cdots , b_R] 
\eea
where $a_r, b_r \in \CC$ with $a_r-b_r \in \ZZ$ for all $r = 1, \cdots , R$.  To a connected decorated graph  $(\Gamma, A, B)$ we associate a complex-valued function  on $\cH$, defined by a multiple Kronecker-Eisenstein sum over  $p_1, \cdots, p_R$ in $\Lambda ' = \Lambda \setminus \{ 0 \}$ where $\Lambda= \ZZ + \tau \ZZ$ for $\tau \in \cH$,  
\bea
\label{2a1}
\cC_\Gamma \left [ \begin{matrix} A \cr B \cr \end{matrix} \right ]  (\tau) 
=  
\sum_{p_1,\dots,p_R \in \Lambda '}  
\,   \prod_{v =1}^V \delta \left ( \sum_{s =1}^R \Gamma _{v \, s} \, p_s \right )
~ \prod_{r =1}^R  { \tau_2^{\half a_r+\half b_r} \over \pi^{\half a_r+\half b_r} (p_r) ^{a_r} ~ (\bar p _r) ^{b_r} }
\eea 
 The Kronecker $\delta$ equals 1 when its argument vanishes and  0  otherwise.    Absolute convergence of the sums in (\ref{2a1}) is ensured by a system of inequalities on the combinations $\Re (a_r+b_r)$, beyond which the functions of (\ref{2a1}) may be defined by analytic continuation in $a_r+ b_r$. 

\sm

The function $\cC_\Gamma$ defined on $\cH$ in (\ref{2a1}) vanishes whenever  $\Gamma$ becomes disconnected upon severing a single edge, and factorizes whenever $\Gamma$ is the union of two subgraphs whose intersection consists of a single vertex. Furthermore, $\cC_\Gamma$ is invariant under the action of $SL(2,\ZZ)$ on $\tau$ if and only if the sums of the $A$ and $B$ exponents are equal to one another, 
\bea
\label{weight}
w = a_1+ \cdots + a_R = b_1+ \cdots + b_R
\eea
The number $w$ is  the {\sl weight}, while $R-V+1$ is the {\sl number of loops} of the modular graph function $\cC_\Gamma$. Bivalent vertices may be eliminated by suitably increasing the values of the exponents $a_r,b_r$.  Henceforth, we shall assume that $\Gamma$ remains connected upon removing any single edge or vertex and that the equality of the sums of $A$ and $B$ exponents in (\ref{weight}) holds.

\sm

It was shown in \cite{DHoker:2015wxz} that the behavior near the cusp of an arbitrary modular graph function of weight $w \geq 2$, whose exponents satisfy (\ref{weight}), is governed by a Laurent polynomial in $\tau_2$ of degree $(w,1-w)$ up to exponentially suppressed terms, 
\bea
\label{Laurent}
\cC_\Gamma \left [ \begin{matrix} A \cr B \cr \end{matrix} \right ]  (\tau)  
= \sum _{k=1-w} ^w \cL_k  (4 \pi \tau_2)^k
+ \cO(e^{-2 \pi \tau_2}) 
\eea
where $\cL_k$ are constants.
Its Poincar\'e series  may be obtained by the arguments used in \cite{DHoker:2019txf} for the more restricted class of dihedral modular graph functions, namely for graphs containing two vertices whose valence is larger than 2. The Poincar\'e series of an arbitrary  modular graph function of weight $w$ with respect to the coset $\Gamma _\infty \backslash PSL(2,\ZZ)$ is given by,
\bea
 \cC_\Gamma \left [ \begin{matrix} A \cr B \cr \end{matrix} \right ]  (\tau)  
 = \sum _{g \, \in \, \Gamma _\infty \backslash PSL(2,\ZZ)} \Lambda _\Gamma \left [ \begin{matrix} A \cr B \cr \end{matrix} \right ]  (g \tau)
 \eea
where $\Gamma _\infty$ is the Borel subgroup of translations of $\tau$, and the seed function $\Lambda_\Gamma$ is given by, 
 \bea
 \label{Poincare}
 \Lambda _\Gamma \left [ \begin{matrix} A \cr B \cr \end{matrix} \right ]  (\tau)
=
{ \tau_2^w \over \pi^w} \sum_{n \not=0}  \, \sum_{p_1,\dots,p_R \in \Lambda '}  
\!\! \delta(p_R-n)  \prod_{v =1}^V \delta \left ( \sum_{s =1}^R \Gamma _{v \, s} \, p_s \right )
~ \prod_{r =1}^R {1 \over  (p_r) ^{a_r} ~ (\bar p _r) ^{b_r} }
\eea
Using the same methods as were used in \cite{DHoker:2019txf} for two-loop modular graph functions, one  shows that the Poincar\'e series in the general case is absolutely convergent whenever the original Kronecker-Eisenstein series is absolutely convergent. We shall next provide more explicit formulas in the cases of one-loop and two-loop modular graph functions.

\subsection{One-loop modular graph functions: Eisenstein series}

A connected one-loop modular graph function of weight $w$ reduces to an Eisenstein series, 
\bea
E_w (\tau) = \sum _{ p \in \Lambda'} { \tau_2^w \over \pi^w |p|^{2w}}
\eea
which satisfies the differential equation,
\bea
\label{diffE}
\Delta E_w = w(w-1) E_w 
\hskip 1in
\Delta = 4 \tau_2^2 \p_{\bar \tau} \p_\tau
\eea
The Fourier series representation of $E_w$ in the variable $\tau_1$ is given by,
\bea
\label{Fourier}
E_w(\tau) = c_w \tau_2^w + \tilde c_w \tau_2 ^{1-w} 
+ { 8 \tau_2 ^\half \over \Gamma (w)} \sum_{N=1}^\infty N^{w-\half} \sigma _{1-2w} (N) 
K_{w-\half} (2 \pi N \tau_2)  \cos (2 \pi N \tau_1) 
\eea
Here, $K_s$ is the modified Bessel function, and we have used the following notations, 
\bea
c_w = { 2 \zeta (2w) \over \pi^w} 
\hskip 0.7in 
\tilde c_w = { 2 \Gamma (w-\half) \zeta (2w-1) \over \Gamma (w) \pi^{w-\half}}
\hskip 0.7in
\sigma_s(N) = \sum _{0< d |N} d^s
\eea 
where $\zeta$ is the Riemann zeta-function. The Poincar\'e series of $E_w$,
\bea
\label{seedE}
E_w (\tau) =  \sum _{g \, \in \, \Gamma _\infty \backslash PSL(2,\ZZ)} c_w (\Im g \tau)^{w}
\eea
is absolutely convergent throughout $\cH$ for $\Re(w) >1$.

\subsection{Two-loop modular graph functions}

The  Kronecker-Eisenstein sum for a two-loop modular graph function with non-negative integer exponents $a_r, b_r$ and weight $w=a_1+a_2+a_3=b_1+b_2+b_3$ simplifies as follows,
\bea
\label{4a1}
\cC \left [ \begin{matrix} a_1 & a_2 & a_3 \cr b_1 & b_2 & b_3 \cr \end{matrix} \right ] (\tau)
= 
\sum _{p_1, \, p_2, \, p_3 \in \Lambda '} \,
{ \tau_2 ^w \, \delta _{p_1+p_2+p_3,0} \over  \pi^w p_1^{a_1} \, p_2^{a_2} \, p_3^{a_3} \, 
\bar p_1^{b_1} \, \bar p_2^{b_2} \, \bar p_3^{b_3} }
\eea 
Every two-loop modular graph function  admits a linear decomposition onto a subset of modular graph functions of a simplified type, 
 \bea
 \cC_{u,v;w}(\tau) =  \cC  \left [ \begin{matrix} u & 0 & w-u \cr 0 & v & w-v \cr \end{matrix} \right ] (\tau)
 \eea
The decomposition was proven in Proposition 2.1 of \cite{DHoker:2019txf} and is given by, 
\bea
\label{thm1}
\cC \left [ \begin{matrix} a_1 & a_2 & a_3 \cr b_1 & b_2 & b_3 \cr \end{matrix} \right ]
& = & 
 \sum _{u=1}^{a_1} \,  \sum _{v=w-b_3}^{w-1}
(-)^{a_3+b_2+u+v+w} \tbinom{w-u-1-a_3}{a_2-1}  \tbinom{v-1-b_2}{b_1-1} \,  \cC_{u,v;w} + (1 \leftrightarrow 2)
 \no \\ && 
+ \sum _{u=w-a_1}^{w-1} \,  \sum _{v=w-b_1}^{w-1} 
 (-)^{a_3+b_2+u+v} \tbinom{u-1-a_3}{a_2-1}   \tbinom{v-1-b_2}{b_3-1} \,   \cC_{u,v;w} + (1 \leftrightarrow 2)
\qquad 
\eea
 where the terms $(1 \leftrightarrow 2)$ are obtained by swapping the pairs $(a_1, b_1)$ with $(a_2, b_2)$ leaving $(a_3, b_3)$ unchanged. Therefore, the Fourier series, Poincar\'e series, and integral of an arbitrary two-loop modular graph function may be expressed in terms of the functions $  \cC_{u,v;w}$.
 
 \sm
 
The Fourier series representation of $\cC_{u,v;w}$ in the variable $\tau_1$ was obtained explicitly 
in \cite{DHoker:2019txf}. The Laurent polynomial part of the constant Fourier modes is given as follows, 
\bea
\label{LaurentC}
\cC_{u,v;w}(\tau) = \ell_w (4 \pi \tau_2)^w  + \sum _{k=0}^{w-2} \ell_{w-2k-3} { \zeta (2k+3) \over (4 \pi \tau_2)^{-w+2k+3}} + { \ell_{2-w} \over ( 4 \pi \tau_2)^{w-2}} + \cO(e^{-4 \pi \tau_2})
\eea
where $\ell_w, \,  \ell_{w-2k-3} \in \QQ$,  the coefficients $\ell_{w-2k-3}$ are given by,
\bea
\label{ell-k}
\ell_{w-2k-3} &=& {(-)^k \zeta(2w-2k-4) \over (2 \pi)^{2w-2k-4}} \bigg ( 
4     (-)^{v+w}   \tbinom{2k+2}{w-u-1}  \tbinom{u+v-2w+2k+3}{u-2w+2k+4} 
\no \\ && \hskip 1.4in
+ 2 \tbinom{2k+2}{-u-v+2k+4} \tbinom{u+v-2}{v-1}  \bigg )  +(u \leftrightarrow v)
\eea
and $\ell_{2-w}$ is a bilinear in odd zeta-values with rational coefficients. The explicit expressions for $\ell_w$ and $\ell_{2-w}$, the non-constant Fourier modes, and the exponentially suppressed part of the constant Fourier mode  were also obtained  in \cite{DHoker:2019txf} but will not be needed here.
 
 \sm
 
The Poincar\'e series of $\cC_{u,v;w}$ with respect to the coset $\Gamma _\infty \backslash PSL(2,\ZZ)$ is given by,
\bea
\label{PoincC}
\cC_{u,v;w}(\tau) = \sum _{g \, \in  \, \Gamma _\infty \backslash PSL(2,\ZZ)} \Lambda _{u,v;w}(g \tau)
\eea
The seed function $ \Lambda _{u,v;w}$ was computed in \cite{DHoker:2019txf} and is given by,
\bea
\label{seedL}
 \Lambda _{u,v;w} (\tau) = \ell _w ( 4 \pi \tau_2)^w + \sum_{m,n \not=0} \, \sum_{\mu \in \ZZ }
 {(-)^v \,  \tau_2^w \over \pi^w n^{2w-u-v} (m\tau+\mu)^u (m \bar \tau + \mu +n)^v}
 \eea
The expressions for the seed function $ \Lambda _{u,v;w} (\tau) $ in terms of elementary functions, given in \cite{DHoker:2019txf}, will not be needed here.

\section{Integrals of modular graph functions}
\setcounter{equation}{0}
\label{sec:3}

 The purpose of this paper is to integrate modular graph functions over $\cM_L$ for finite $L>1$,
 \bea
 \cS_\Gamma \left [ \begin{matrix} A \cr B \cr \end{matrix} \right ]  (L)
 = \int _{\cM_L}  {d^2 \tau \over \tau_2^2} \, \cC_\Gamma \left [ \begin{matrix} A \cr B \cr \end{matrix} \right ]  (\tau)
 \eea
 The $L$-dependence of $\cS_\Gamma$ is given by the integral of the constant Fourier mode of $\cC_\Gamma$, 
  \bea
 \cS_\Gamma \left [ \begin{matrix} A \cr B \cr \end{matrix} \right ]  (L) = 
  \cS_\Gamma \left [ \begin{matrix} A \cr B \cr \end{matrix} \right ]  (1)
 + \int _1^L  {d \tau_2 \over \tau_2^2} \, \int _0 ^1 d \tau_1 \, 
 \cC_\Gamma \left [ \begin{matrix} A \cr B \cr \end{matrix} \right ]  (\tau)
 \eea
thereby determining the full integral $\cS_\Gamma$ in terms of the constant Fourier mode  up to an additive constant.  As $L \to \infty$, the behavior of $\cS_\Gamma$ is determined by the Laurent polynomial part of the constant Fourier mode, up to this additive integration constant and up to exponentially suppressed corrections, 
\bea
\cS_\Gamma (L) = \cL_c + \cL_1 \ln L + \sum _{{k =1-w \atop k \not=1}}^w  { \cL_k \over k-1} L^{k-1} + \cO(e^{-4 \pi L})
\eea
For string theory applications it is the constant  $\cL_c$, which is not determined by the Laurent polynomial, in which we are most interested. In the sequel we shall begin by reviewing the results of Zagier for the case when the modular function is linear or bilinear in the Eisenstein series. Next, we shall present new results for the case where $\cC_\Gamma$ is an arbitrary two-loop modular graph function, and obtain the integrals  using the Poincar\'e series of $\cC_\Gamma$.

\subsection{Integrals of Eisenstein series and their products}
\label{sec:31}

The integral of $E_w$ over the domain $\cM_L$ may be simplified by using the differential equation (\ref{diffE}) and Green's theorem to recast it as an integral over the boundary of $ \cM_L$, 
\bea
w(w-1) \int _{\cM_L} { d^2 \tau \over \tau_2^2} E_w =  \int _{\cM_L} { d^2 \tau \over \tau_2^2} \Delta E_w =  \int _0 ^1 d \tau_1 \p_{\tau_2} E_w \Big |_{\tau_2=L}
\eea
Using the Fourier series for $E_w$ of (\ref{Fourier})  in the right most integral, we see that only the constant Fourier mode contributes and the integral may be evaluated exactly,
\bea
\label{Ewint}
\int _{\cM_L} { d^2 \tau \over \tau_2^2} E_w = { c_w \, L^{w-1} \over w-1} - { \tilde c _w \, L^{-w} \over w}
\eea
Away from $w=0$ and $w=1$ the integral is an analytic function of $w$. For $0 < \Re(w) < 1$, the right side tends to zero as $L \to \infty$, resulting in the vanishing of the integral of $E_w$ over the full fundamental domain $\cM$,\footnote{ It has been assumed in  \cite{Zagier} that analytic continuation in $w$ may be used to set the integral over $\cM$ to zero throughout $w \in \CC$,  but we shall not need or use this assumption here.}
\bea
\label{Eis-int}
\int _\cM { d^2 \tau \over \tau_2^2} E_w = \lim _{L \to \infty} \int _{\cM_L} { d^2 \tau \over \tau_2^2} E_w =0 
\eea
The integral of a product of two Eisenstein series may be simplified in a similar manner, 
\bea
\Big ( w'(w'-1) - w(w-1) \Big ) \int _{\cM_L} { d^2 \tau \over \tau_2^2} E_w E_{w'} 
& = & \int _{\cM_L} { d^2 \tau \over \tau_2^2} \Big ( E_{w} \Delta E_{w'} - E_{w'} \Delta E_w \Big )
\\
& = &
\int _0^1 d \tau_1 \Big ( E_{w} \p_{\tau_2}  E_{w'} - E_{w'} \p_{\tau_2}  E_w \Big ) \Big |_{\tau_2 =L}
\no
\eea
In this case the non-constant Fourier modes of $E_w$ and $ E_{w'}$ will contribute to the boundary term as well, but their effect will be exponentially suppressed for large $L$, and we obtain an equation which is equivalent to the Maass-Selberg relation \cite{Zagier}, 
\bea
\label{EwEwp}
\int _{\cM_L} \!\! { d^2 \tau \over \tau_2^2} E_w E_{w'} 
\! = \! 
{c_w c_{w'} L^{w+w'-1} \over w+w'-1} \! +  \! { \tilde c_w \tilde c_{w'} L^{1-w-w'} \over 1-w-w'} 
\! + \! { c_w \tilde c_{w'} L^{w-w'} \over w-w'} \! +  \! { \tilde c_w c_{w'} L^{w'-w} \over w'-w} 
\! +  \! \cO(e^{-4 \pi L})
\quad
\eea
In this case the region in $w,w'$ where the right side tends to zero as $L \to \infty$ is empty. However, if $\Re(w-w') > 0$ and $\Re(w+w')>1$  we may subtract a suitably chosen linear combination of $E_{w+w'}$ and $E_{w-w'+1}$ to cancel the growing terms  in the limit $L \to \infty$, leading to the following convergent integral \cite{Zagier}, 
\bea
\int _{\cM} { d^2 \tau \over \tau_2^2} \left ( E_w E_{w'} 
- { c_w c_{w'} \over c_{w+w'}} \, E_{w+w'} 
-{ c_w \tilde c_{w'} \over c_{w-w'+1}} \, E_{w-w'+1} \right ) =0
\eea
The integral of $E_w^2$ may be obtained by taking the limit as $w' \to w$ of (\ref{EwEwp}) and we find, 
\bea
\label{EEint}
\lim _{L \to \infty} \left ( \int _{\cM_L} { d^2 \tau \over \tau_2^2} \left \{ E_w^2 
- { c_w^2  \over c_{2w}} \, E_{2w} \right \} - 2 c_w \tilde c_w \ln L \right )  
= \tilde c_w  \, { \p c_w \over \p w}   -  c_w \, {\p \tilde c_w \over \p w} 
\eea
This result was obtained  in Appendix A of \cite{Green:2008uj}.

\subsection{Integral of two-loop modular graph functions of odd weight}

When the weight $w \geq 3 $ of the modular graph function $\cC_{u,v;w}$  is odd, the terms in (\ref{LaurentC})  that grow with $\tau_2$ near the cusp are the leading term proportional to $\tau_2^w$, as well as the sub-leading terms with $0 \leq 2k \leq w-5$. Therefore, the integral of $\cC_{u,v;w}$ over $\cM$ diverges near the cusp.  We shall define a convergent integral by subtracting from $\cC_{u,v;w}$ a linear combination of Eisenstein series $E_s$ for integer  with $2\leq s \leq w$, chosen so as to cancel the growing behavior near the cusp. Specifically, we define the following modular function, 
\bea
\label{hatC}
\hat \cC _{u,v;w} = \cC_{u,v;w}  - \ell_w { (2 \pi)^{2w} \over 2 \zeta(2w)} E_w
- \sum _{k=0}^{ { w-5 \over 2} } \ell_{w-2k-3}  
{ (2\pi)^{2w-4k-6}  \zeta (2k+3)  \over 2 \zeta (2w-4k-6)} E_{w-2k-3}
\eea
For the range of $k$ in the above sum, the expression for the coefficients $\ell_{w-2k-3}$ given in (\ref{ell-k}) simplifies, since $0 \leq 2k \leq w-5$ and $u < w$ imply $u-2w+2k+4 \leq u-w-1 <0$ and, as  a result, the first term in (\ref{ell-k}) for $\ell_{w-2k-3}$ vanishes. The remaining term is manifestly symmetric under $u \leftrightarrow v$ and may be expressed as follows, 
\bea
\label{simp-ell}
\ell_{w-2k-3} = 4 (-)^k   \binom{2k+2}{u+v-2} \binom{u+v-2}{u-1}  { \zeta(2w-2k-4) \over (2\pi)^{2w-2k-4}} 
\eea
where the first binomial coefficient vanishes for $2k+4<u+v$.  By construction, the function $\hat \cC_{u,v;w}$ tends to a constant at the cusp, and is  integrable on the full fundamental domain $\cM$. The integral is given by the following theorem.

{\thm
\label{thm1}
{\sl For $w=2\kappa+1\geq 3$ an odd integer and positive integers $u,v$ satisfying $u+v\geq 3$ and $u,v\leq w-1$,  the integral of  $\hat \cC_{u,v;w}$ over $\cM$,
\bea
\label{ChatC}
\hat \cS_{u,v;w} 
=  \int _\cM {d^2 \tau \over \tau_2^2} \, \hat \cC_{u,v;w} (\tau) 
\eea
evaluates to, 
\bea
\label{cS}
\hat \cS_{u,v;w} = 2 \pi \zeta(2\kappa+1)  \binom{2\kappa-1}{2\kappa+1-u-v} \binom{u+v-2}{u-1} {B_{2\kappa} \over (2\kappa)!}
\eea
where $B_{2\kappa}$ are the Bernoulli numbers.  The integral of $\cC_{u,v;w}$ over $\cM_L$ follows by combining (\ref{Ewint}) and (\ref{hatC}), up to contributions which vanish for $L \to \infty$.}}

\sm

Note that the first binomial coefficient vanishes when $w< u+v$, and that $\hat \cS_{u,v;w}$ is given by a rational number times an odd zeta-value, and is therefore a single-valued zeta-value.

\sm

To prove the  theorem, namely equation (\ref{cS}), we evaluate the integral of $\hat \cC_{u,v;w}$ using the Poincar\'e series for $\hat \cC_{u,v;w}$, which may be obtained by combining the Poincar\'e series of $\cC_{u,v;w}$  and $E_s$ using (\ref{hatC}). The result is the following Poincar\'e series,   which is absolutely convergent \cite{DHoker:2019txf}, 
\bea
\hat \cC_{u,v;w}(\tau) = \sum _{g \, \in \, \Gamma _\infty \backslash PSL(2,\ZZ)} \hat \Lambda _{u,v;w} (g\tau)
\eea
The seed function $\hat \Lambda _{u,v;w}$ is obtained by combining the seed function of $\cC_{u,v;w}$ in (\ref{seedL})  with the seed function for $E_s$ in (\ref{seedE}) using again (\ref{hatC}), and we have, 
\bea
\hat \Lambda _{u,v;w}(\tau) & = &  \sum _{m,n \not=0} \sum _{\mu \in \ZZ} 
{ (-)^v \, \tau_2^w \over \pi^w n^{2w-u-v} (m\tau+\mu)^u (m \bar \tau + \mu +n)^v}
\\ &&
-  \sum _{k=0}^{ { w-5 \over 2} }  4(-)^k  \zeta (2k+3)   \zeta(2w-2k-4)    
\tbinom{2k+2}{-u-v+2k+4} \tbinom{u+v-2}{v-1}  { (2 \tau_2)^{w-2k-3}  \over (2\pi)^{w-1}}
\no
\eea
Using the standard unfolding trick for the integral of a Poincar\'e series, we have,
\bea
\hat \cS_{u,v;w} = \int _0^\infty {d \tau_2 \over \tau_2^2} \int _0 ^1 d \tau_1 \, \hat \Lambda _{u,v;w}(\tau)
\eea
The sum over the integer $\mu$ in the first line may be carried out by partial fraction decomposition and use of the following standard summation formula,
\bea
\label{Lif}
\sum_{\mu \in \ZZ} { 1 \over (z+\mu)^{k+1}} 
= i \pi  \, {(-)^k \over k!} { d^k \over dz ^k } \left ( { 1 + e^{2 \pi i z} \over 1-e^{2 \pi i z}} \right )
\eea
The integral over $\tau_1$ of the partial fractions projects onto the constant Fourier mode, which contributes only for $k=0$, and we obtain, 
\bea
\label{Int1}
\int _0^1 d \tau_1 \sum _{\mu \in \ZZ} { 1 \over (m\tau + \mu)^u (m \bar \tau + \mu +n )^v} 
=
-2 \pi i { (-)^v \ep (m) \binom{u+v-2}{u-1}  \over (2 i m \tau_2-n)^{u+v-1}}
\eea
where $\ep(m)$ evaluates to 1 for $m>0$, $-1$ for $m<0$, and 0 for $m=0$.
As a result, we find, 
\bea
\hat \cS_{u,v;w} & = & { \binom{u+v-2}{u-1}  \over \pi^{w-1}} \int _0^\infty \!\! d\tau_2  \Bigg [
\sum _{m,n=1}^\infty 
{  \tau_2^{w-2}   \over  n^{2w-u-v}} \left (  { - 4 i  \over (2 i m \tau_2-n)^{u+v-1}} -  { - 4 i  \over (-2 i m \tau_2-n)^{u+v-1}} \right )
\no \\ && \hskip 0.6in
- \sum _{k=0}^{ { w-5 \over 2} } 4(-)^k \zeta (2k+3)   \zeta(2w-2k-4)    \tbinom{2k+2}{-u-v+2k+4}  
{  \tau_2^{w-2k-5}  \over 2^{2k+2}}  \Bigg ]
\eea
To obtain the first term we have used the invariance under $(m,n) \to (-m,-n)$ to restrict the sum over $m$ to $m >0$ upon including a factor of 2, and then decomposed the $n$-sum into its contributions from positive and negative $n$. The presence of the subtraction terms on the second line above, which resulted from the subtraction of Eisenstein series, guarantees the absolute convergence of the integral. 

\sm

One would like to rescale $\tau_2$ in the first line of the integrand to factor out the $m$ and $n$-dependences. But the integral over each line separately  does not converge, so we need to rescale all terms in the integrand simultaneously. To do so, we substitute for $\zeta (2k+3)$ and $\zeta(2w-2k-4)$ their infinite series representations respectively  in the summation variables $m$ and $n$, and then change variables  from $\tau_2 $ to $x=2m\tau_2/n$.  Doing so, all dependence on $m$ and $n$ factors out and may be summed  in terms of Riemann zeta-values. 
Next, we use the fact that, for $w$ odd, the integrand is even in $x$ in order to extend the integration range to all of $\RR$, upon including a factor of $\thalf$, and we obtain,  
\bea
\hat \cS_{u,v;w} & = & { \binom{u+v-2}{u-1} \zeta(w-1) \zeta(w) \over (2\pi)^{w-1}} 
\int _{-\infty} ^\infty dx \, \bigg [
 { - 2 i x^{w-2} \over ( i x-1)^{u+v-1}} +  {  2 i x^{w-2} \over (- i x-1)^{u+v-1}}
\no \\ && \hskip 2.2in
- \sum _{k=0}^{ { w-5 \over 2} }  4 (-)^k \tbinom{2k+2}{-u-v+2k+4}     x^{w-2k-5}    \bigg ]
\eea
The subtraction terms guarantee that the integrand tends to zero as $1/x^2$ for large $x$, so we may close the contour of integration in the upper half plane and evaluate the integral via the residue of the integrand  at the single pole at $x=i$. Finally, the even zeta-values may be converted into Bernoulli numbers using the relation,
\bea
\zeta(2\kappa) = \half (-)^{\kappa+1} (2 \pi)^{2\kappa} \, { B_{2\kappa} \over (2 \pi)^{2\kappa}}
\eea
which gives the result announced in (\ref{cS}) of Theorem (\ref{thm1}), thereby completing its proof.

\sm

We note that the value of the integral $\hat \cS_{u,v;w}$ is closely related to the term of order $\tau_2^0$ in the Laurent polynomial of $\cC_{u,v;w}$, which is given by $\ell_0 \zeta (w)$. This term corresponds to $2k=w-3$ in the sum of (\ref{LaurentC}) and, since we have $u-2w+2k+4=u-w+1<0$, the first term in the general expression for $\ell_{w-2k-3}$ does not contribute. As a result, we  find, 
 \bea
 \hat \cS_{u,v;w} = \pi \, { w-u-v+1 \over w-1} \, \ell_0 \, \zeta (w)
 \eea
 where we recall that $\ell_0$ is a rational number dependent on $u,v,w$.

\subsection{Integral of two-loop modular graph functions of even weight}
 
 The case of even $w$ presents a complication, which was absent for odd $w$, due to the presence in the Laurent polynomial of $\cC_{u,v;w}$ of the term linear in $\tau_2$. On the one hand this term will lead to a logarithmic divergence of the integral of $\cC_{u,v;w}$ at the cusp. On the other hand, it cannot be regularized by subtracting an Eisenstein series since $E_s$ has a pole in $s$ at $s=1$. We shall be led to subtracting a more complicated modular graph functions whose integral can be evaluated with the use of the differential equation it satisfies, along the lines of the integrals carried out in subsection \ref{sec:31} for products of Eisenstein series. The result is given by the following theorem.
 
 {\thm
 \label{thm2}
{\sl For $w=2\kappa \geq 4$ an even integer and positive integers $u,$ satisfying $u+v \geq 3$ and $u,v \leq w-1$, we define the modular function  $\tilde \cC_{u,v;w}$ in terms of $\cC_{u,v;w}$ as follows,
\bea
\tilde \cC_{u,v;w} & = & \cC_{u,v;w} - \ell_w { (2\pi)^{2w} E_w \over 2 \zeta (2w)} 
   - \sum _{k=0} ^{{w-6 \over 2}} \ell_{w-2k-3}  { (2 \pi )^{2w-4k-6} \, \zeta (2k+3)  \over 2 \zeta (2w-4k-6)}
   E_{w-2k-3}
\eea
The coefficients  $\ell_{w-2k-3}$ were given in (\ref{simp-ell}) for the range $0\leq 2k\leq w-6$ needed here. Its regularized integral, which is defined by, 
\bea
\label{int-lim}
\tilde \cS_{u,v;w} = \lim _{L \to \infty} \left ( \int _{\cM_L} { d^2 \tau \over \tau_2^2} \, \tilde \cC_{u,v;w} (\tau) - 4 \pi \ell_1 \zeta (w-1) \ln (2L) \right )
\eea
evaluates as follows, 
\bea
\label{EvenC}
\tilde \cS_{u,v;w} 
& =  &  \pi \zeta(2\kappa-1) (-)^{\kappa+1} { B_{2\kappa} \over (2\kappa )!}  \binom{u+v-2}{u-1}  \cG_{u,v;w }  
\no \\ &&
+ 4 \pi \ell_1 \zeta (2\kappa -1)  \left ( { \zeta '(2\kappa ) \over \zeta (2\kappa )} - {\zeta'(2\kappa -1) \over \zeta (2\kappa -1)} \right )
\eea
The function   $\cG_{u,v;w}$ takes  rational values and may be expressed as a linear combination of finite harmonic sums with integer coefficients whose detailed form will be given in (\ref{g}).  }}

\sm

Before proving the theorem, a comment on the transcendentality weight of the result is in order. The first term in $\hat \cS_{u,v;w}$ is a rational number times $\pi \zeta (2\kappa-1)$ times $\cG_{u,v;w}$, which is a finite sum with integer coefficients of finite harmonic sums. It is argued in \cite{DG} that such finite harmonic sums should be assigned transcendental weight one. Similarly, it is argued in \cite{DG} that the logarithmic derivatives that enter $\hat \cS_{u,v;w}$ should also be assigned transcendental weight one. Therefore, both terms in $\hat \cS_{u,v;w}$ have transcendental weight $w+1=2\kappa+1$.

\sm

To prove Theorem \ref{thm2} we first verify  the existence of a finite limit in (\ref{int-lim}) by observing that the function $\tilde \cC_{u,v;w}$ grows  as $4 \pi \tau_2 \ell_1 \zeta(2\kappa-1)$ near the cusp, where $\ell_1$ is,
\bea
\ell_1= -2 \binom{2 \kappa -2}{u+v-2} \binom{u+v-2}{u-1} { B_{2\kappa}  \over ( 2 \kappa) !}
\eea
so that its integral over $\cM_L$ grows as $4 \pi  \ell_1 \zeta(2\kappa-1) \ln(L) $.  

\sm

We cannot directly apply the unfolding trick to the computation of the integral of $\tilde \cC_{u,v;w}$ using its Poincar\'e series, since the simplest unfolding is valid for convergent integrals on the entire fundamental domain $\cM$. To obtain a convergent integral we must make a further subtraction to eliminate the term linear in $\tau_2$. One option would be to subtract a term proportional to $E_s^2$ with $s=w/2$. The problem with this option  is that its Poincar\'e series with seed function proportional to $\tau_2 ^s E_s$ fails to be integrable due to a divergence as $\tau_2 \to 0$.

\subsubsection{Subtraction of the term linear in $\tau_2$}

Instead we shall choose to subtract a term proportional to a fixed modular graph function, 
 \bea
 C_{2,1,1} -{ 2 \over 3} E_4 = \cC_{2,2;4} - { 3 \over 2} ( \cC_{2,3;4} + \cC_{3,2;4}) - { 2 \over 3 } E_4
 \eea
 whose behavior at the cusp may be deduced from (\ref{LaurentC}) and  is given by,
  \bea
 C_{2,1,1} - { 2 \over 3} E_4 =  { \pi \zeta (3) \tau_2 \over 45} + \cO(\tau_2^{-1})
 \eea
 Using this result, we define the modular function $\hat \cC_{u,v;w}$ by,
  \bea
  \hat \cC_{u,v;w}  =  \tilde \cC_{u,v;w} 
  - { 180 \over \zeta(3)} \, \ell_1 \, \zeta (w-1) \, \left ( C_{2,1,1} -{2 \over 3} E_4 \right )
   \eea 
By construction, $ \hat \cC_{u,v;w}$ tends to zero at the cusp and is  integrable on $\cM$. Its integral,  
 \bea
  \hat \cS_{u,v;w} =  \int _{\cM} { d^2 \tau \over \tau_2^2} \, \hat \cC_{u,v;w} (\tau)
 \eea
 may be simply related to the integral $\tilde \cS_{u,v;w}$ we seek by, 
 \bea
 \label{tildeS}
 \tilde S_{u,v;w} = \hat \cS_{u,v;w} + 4 \pi \ell_1 \zeta (w-1) \lim _{L \to \infty} \left ( 
 { 45 \over \pi \zeta (3)} \int _{\cM_L} { d^2 \tau \over \tau_2^2}   \left \{ C_{2,1,1} -{2 \over 3} E_4 \right \} - \ln (2L) \right )
 \eea
 The contribution under the limit may be computed using the differential equation, 
 \bea
 (\Delta -2) C_{2,1,1} = 9 E_4 - E_2^2
 \eea
 to express $C_{2,1,1}$ as a linear combination of $\Delta C_{2,1,1}$, $E_4$ and $E_2^2$,
 \bea
 C_{2,1,1} -{ 2 \over 3} E_4 = \half \Delta \left ( C_{2,1,1} -{ 2 \over 3} E_4 \right ) +\half \left ( E_2^2 - { 7 \over 3} E_4 \right )
 \eea
The integral of the first term on the right side  may be evaluated using Green's theorem, while the integral of the second term may be evaluated using (\ref{EEint}) for the case $w=w'=2$. Putting all together, we have,
\bea
\label{intC211}
 \lim _{L \to \infty} \left ( 
 { 45 \over \pi \zeta (3)} \int _{\cM_L} { d^2 \tau \over \tau_2^2}   \left \{ C_{2,1,1} -{2 \over 3} E_4 \right \} - \ln (2L) \right )
 = { \zeta'(4) \over \zeta (4)} - { \zeta '(3) \over \zeta(3)}
 \eea
 
 \subsubsection{Unfolding the integral of the Poincar\'e series of $\hat \cC_{u,v;w}$}
 
It remains to evaluate the integral $\hat S_{u,v;w}$, which we do using its Poincar\'e series and the standard unfolding trick on the fundamental domain $\cM$.  The Poincar\'e series is given by, 
  \bea
\hat \cC_{u,v;w}(\tau) = \sum _{g \, \in \, \Gamma _\infty \backslash PSL(2,\ZZ)} \hat \Lambda _{u,v;w} (g\tau)
\eea
in terms of which the integral is given by,
\bea
\label{SLambda}
\hat \cS_{u,v;w} = \int_0^\infty { d\tau_2 \over \tau_2^2} \int _0^1 d \tau_1 \, \hat \Lambda _{u,v;w} (\tau)
\eea
The seed function is obtained from the seed function $\Lambda _{u,v;w} $ in (\ref{seedL}), the seed function $c_w \tau_2^w$ for $E_w$, and the seed function for $C_{2,1,1}$, expressed in terms of the seed functions for $\cC_{2,2;4}$, $\cC_{2,3;4}$ and $\cC_{3,2;4}$, and we have, 
\bea
\hat \Lambda _{u,v;w}(\tau) & = &  \sum _{m,n \not=0}  \sum _{\mu \in \ZZ} 
{ (-)^v \, \tau_2^w n^{u+v} \over \pi^w n^{2w} (m\tau+\mu)^u (m \bar \tau + \mu +n)^v}
\\ &&
- \sum _{k=0}^{ { w-6 \over 2} } 4 (-)^k \zeta (2k+3)   \zeta(2w-2k-4)   
\tbinom{2k+2}{-u-v+2k+4} \tbinom{u+v-2}{v-1}  { (2 \tau_2)^{w-2k-3}  \over (2\pi)^{w-1}}
\no \\ &&
- { 180 \, \tau_2^4 \over \zeta(3) \, \pi^4 } \, \zeta(w-1) \,  \ell_1 \sum _{m,n \not=0}  \sum _{\mu \in \ZZ} \bigg [ 
{ 1 \over n^4 (m\tau+\mu)^2 (m\bar \tau + \mu+n)^2} 
\no \\ && \hskip 1in
+ { {3 \over 2} \over n^3 (m\tau+\mu)^2 (m\bar \tau+\mu +n)^3} - { {3 \over 2} \over n^3 (m\tau+\mu)^3 (m\bar \tau+\mu +n)^2} \bigg ]
\no \eea
The summation and integration procedures are analogous to those we used in the case of $w$ odd, and we shall be brief on the parts of the calculation that are similar.  We carry out partial fraction decompositions in the variable $\mu$, using the general formula,
\bea
\label{parfrac}
{ 1 \over (\mu +x)^u (\mu +y)^v} = 
\sum _{k=1}^u {(-)^v \binom{u+v-k-1}{u-k}  \over (\mu +x)^k(x-y)^{u+v-k}} 
+ \sum _{k=1}^v { (-)^u \binom{u+v-k-1}{v-k}  \over (\mu +y)^k (y-x)^{u+v-k}}
\eea
and use (\ref{Lif}) to perform the sums over $\mu$ and (\ref{Int1}) to do the integrals over $\tau_1$. Next, we exploit the symmetry of the summand under $(m,n) \to (-m,-n)$ to restrict the sums to $m>0$ upon including a factor of 2 and separate the contributions from positive and negative $n$. Finally, we express $\zeta (2k+3)$ and $\zeta(2w-2k-4)$ as sums respectively over variables $m$ and $n$ as we did in the case of odd $w$. The result of these manipulations is as follows, 
\bea
\int _0^1 d \tau_1 \, \hat \Lambda _{u,v;w}(\tau) & = &  
{ \tbinom{u+v-2}{u-1} \over (2 \pi)^{w-1}} \sum _{m,n=1}^\infty  \Bigg [
{ (2 \tau_2)^w \over  n^{2w-u-v} } \left ( {- 2 i   \over (2 i m \tau_2 -n)^{u+v-1}} + { 2 i   \over (- 2 i m \tau_2 -n)^{u+v-1}} \right ) 
\no \\ &&
- \sum _{k=0}^{ { w-6 \over 2} } 4 (-)^k { 1 \over m^{2k+3} n^{2w-2k-4}}     \tbinom{2k+2}{-u-v+2k+4}   (2 \tau_2)^{w-2k-3} 
\no \\ &&
+  { 180 \,   \tau_2^4  \over \zeta(3) \,  \pi^4 }  \zeta(w-1) \zeta(w) \,  (-)^{{w \over 2}}  \tbinom{w-2}{w-u-v} 
  \bigg (  { 16i  \over n^4 (2im\tau_2-n)^3} 
\no \\ && \hskip 0.4in
+ { 16i \over n^4 (2im\tau_2+n)^3}  + { 72i \over n^3(2 i m \tau_2-n)^4}  -  { 72i  \over n^3(2 i m \tau_2+n)^4}  \bigg )
\Bigg ]
\eea

\subsubsection{Carrying out the integral over $\tau_2$}

In carrying out the remaining integral over $\tau_2$, a crucial subtlety arises which did not occur in the case of odd $w$. The partial fraction decomposition of the integrand now contains an infinite sub-series  which decays to zero as $1/\tau_2$ at the cusp. To interchange the summations in the infinite series and the integration, and to carry out the integrations term by term, we shall first regularize the integral by imposing a cutoff $\tau_2 <L$, and treat the terms in $1/\tau_2$ separately. To do so, we introduce the following notation, 
\bea
\hat \cS_{u,v;w} 
= { \tbinom{u+v-2}{u-1} \over (2 \pi)^{w-1}}  \left ( \cF_{u,v;w}^+ + \cF^-_{u,v;w} \right )
\eea
The contribution $\cF^-_{u,v;w}$ collects all the terms of order $1/\tau_2$  in the partial fraction decomposition of the integrand, and is given by, 
\bea
\cF^-_{u,v;w} & = & 4 (-)^{{w \over 2}} \tbinom{w-2}{w-u-v}  \lim _{L \to \infty} 
\int _0 ^L  d \tau_2 \sum_{m,n=1}^\infty \bigg [
{ 1  \over n^w m^{w-1} } \left ( { 1 \over \tau_2 + {in\over m}} +  { 1 \over \tau_2 - {in\over m}} \right )
\no \\ && \hskip 1.8in
- {  \zeta(w-1) \zeta(w) \over \zeta(3)  \zeta(4) n^4 m^3} \left ( { 1 \over  \tau_2 + {in\over m} }
+ { 1 \over  \tau_2 - {in\over m} } \right ) \bigg ]
\eea
The integrals are readily evaluated. In the limit $L \to \infty$, all dependence on $L$ drops out of the summations over $m,n$. The summations over $m,n$ may be performed in terms of the Riemann zeta function and its first derivative, and we find,  
\bea
\label{fm}
\cF^-_{u,v;w} = 
8 (-)^{{w \over 2}} \tbinom{w-2}{w-u-v}  \zeta(w) \zeta(w-1)  
\left ( { \zeta '(w) \over \zeta(w)} -  { \zeta '(w-1) \over \zeta(w-1)} 
-  { \zeta '(4) \over \zeta(4)}  +  { \zeta '(3) \over \zeta(3)}  \right )
\eea
The contribution $\cF^+_{u,v;w}$ collects all the remaining terms in the integral. Changing variables from $\tau_2$ to $y=2m \tau_2/n$ in this absolutely convergent integral,  the summations over $m$ and $n$ may be performed in terms of Riemann zeta-values. The terms in $1/y$ which remain after this change of variables cancel  in the integrand, and we are left with,
\bea
\label{f}
\cF_{u,v;w}^+ = \zeta(w) \zeta (w-1) \cG_{u,v;w}
\eea
where the function $\cG_{u,v;w}$ is given by the integral, 
\bea
\label{g}
\cG_{u,v;w} & = & \int _0 ^\infty  d y  \Bigg ( 
{ 4 \, (-i )^{u+v} \,  y^{w-2}  \over ( y +i)^{u+v-1}} + { 4 \,  i^{u+v}  \,   y^{w-2}  \over (y -i)^{u+v-1}} 
- \sum _{k=0}^{ { w-6 \over 2} } 8 (-)^k    \tbinom{2k+2}{-u-v+2k+4}   y^{w-2k-5} 
\no \\ && \qquad
-   (-)^{{w \over 2}}  \tbinom{w-2}{w-u-v}   \Big (  
{ 4 y^2 \over  (y+i)^3} + { 4 y^2 \over  (y-i)^3} 
 - { 18i y^2 \over ( y+i)^4}  +  { 18 i y^2  \over (  y-i)^4}  \Big ) \Bigg )
\eea
Assembling all contributions to $\tilde \cS_{u,v;w}$ we observe that the contributions involving $\zeta'(3)$ and $\zeta'(4)$ cancel out between $\cF^-_{u,v;w}$ and the combination of (\ref{tildeS}) and (\ref{intC211}), and we obtain the result announced in Theorem \ref{thm2}, with the expression for $\cG_{u,v;w}$ given by (\ref{g}).

\subsubsection{Evaluating $\cG_{u,v;w}$}

Finally, we evaluate $\cG_{u,v;w}$ by decomposing the  integrand into partial fractions. All polynomially growing or constant contributions cancel by construction. Similarly, all terms that grow as $1/y$ for large $y$ also cancel one another. Integrating the remaining rational functions produces only rational numbers, and we find, 
\bea
\cG_{u,v;w} = 24 (-)^{{w \over 2}} \tbinom{w-2}{u+v-2}
+ 8 (-)^{{w \over 2}} \sum _{\ell=1}^{u+v-2} \tbinom{w-2}{u+v-2-\ell} { (-)^\ell \over \ell}
\eea
Defining harmonic sums by,
\bea
H_\ell = \sum _{m =1}^\ell { 1 \over m} \hskip 1in H_0=0
\eea
the function $\cG_{u,v;w}$ becomes a sum, with integer coefficients, of harmonic sums,
 \bea
\cG_{u,v;w} = 24 (-)^{{w \over 2}} \tbinom{w-2}{u+v-2}H_1
+ 8 (-)^{{w \over 2}} \sum _{\ell=1}^{u+v-2} (-)^\ell \tbinom{w-2}{u+v-2-\ell}   (H_\ell - H_{\ell-1})
\eea

\section{Higher loops: open problems}
\setcounter{equation}{0}
\label{sec:4}

The regularized integrals of various individual modular graph functions at three-loop order and beyond have been evaluated using a combination of techniques, sometimes specific to a given modular graph function under study. General techniques include reducing the loop order of the integrand by the use of the algebraic  identities  and exposing total derivatives by the use of the differential identities satisfied by the  modular graph functions. Although a general algorithm is known to find all algebraic identities at a given weight  \cite{DHoker:2016mwo,DHoker:2016quv}, the procedure for finding the identities is involved, and has been systematically worked out only to weight six included, with some identities known also at weight seven \cite{DHoker:2016quv}. A different, but related,  algorithm based on  iterated Eisenstein integrals has been developed in  \cite{Broedel:2018izr}. An explicit expression for the integral of a modular graph functions of arbitrary loop order and arbitrary weight has not, however, been obtained thus far.

\sm

It is possible, however, to formulate the algorithm by which the integral of a modular graph function of arbitrary loop order and weight can be evaluated using only elementary integrals. We shall now describe the steps required. Consider a modular graph function $\cC_\Gamma$ of weight $w$ given by an absolutely convergent Kronecker-Eisenstein series in (\ref{2a1}), its Laurent polynomial in (\ref{Laurent}), and its absolutely convergent Poincar\'e series in (\ref{Poincare}). Note that the values of $\cL_k$ in the Laurent polynomial of (\ref{Laurent}) are not known for arbitrary loop order and weight, but can be (laboriously) calculated at any order if needed. Following the procedure used at two-loops, we construct a new modular function, 
\bea
\hat \cC_\Gamma \left [ \begin{matrix} A \cr B \cr \end{matrix} \right ]    & = & 
\cC_\Gamma \left [ \begin{matrix} A \cr B \cr \end{matrix} \right ]  
- \sum_{k=2}^w {(4 \pi)^k \cL_k \over c_k} E_k - { 180 \, \cL_1 \over \zeta (3)} \left ( C_{2,1,1} - { 2 \over 3} E_4 \right )
\eea  
By construction, $\hat \cC_\Gamma$ tends to a constant at the cusp and is therefore integrable on $\cM$. 
The integral of  $\hat \cC_\Gamma$ is obtained using the unfolding trick, and may be expressed in terms of the integral of its Poincar\'e seed, 
\bea
 \int _{\cM} { d^2 \tau \over \tau_2^2} \, \hat \cC_\Gamma \left [ \begin{matrix} A \cr B \cr \end{matrix} \right ]   (\tau) 
= \int _0 ^\infty { d \tau_2 \over \tau_2^2} \int _0 ^1 d \tau_1 \, \hat \Lambda_\Gamma \left [ \begin{matrix} A \cr B \cr \end{matrix} \right ]   (\tau) 
\eea
where the seed function $\hat \Lambda _\Gamma$  is given in terms of the seed function $\Lambda _\Gamma$ for $\cC_\Gamma$  by,
\bea
\hat \Lambda_\Gamma \left [ \begin{matrix} A \cr B \cr \end{matrix} \right ]   (\tau) 
& = & \Lambda _\Gamma \left [ \begin{matrix} A \cr B \cr \end{matrix} \right ]  (\tau) 
- \sum_{k=2}^w  \cL_k (4 \pi \tau_2)^k 
\\ &&
- { 90 \, \tau_2^4  \,\cL_1 \over \zeta(3) \, \pi^4 } \sum _{m,n \not=0}  \sum _{\mu \in \ZZ} \bigg [ 
{ 2 \over n^4 (m\tau+\mu)^2 (m\bar \tau + \mu+n)^2} 
\no \\ && \hskip 0.5in
+ { 3 \over n^3 (m\tau+\mu)^2 (m\bar \tau+\mu +n)^3} - { 3 \over n^3 (m\tau+\mu)^3 (m\bar \tau+\mu +n)^2} \bigg ]
\no
\eea 
By construction, the integral of the Poincar\'e seed $\hat \Lambda _\Gamma$ converges absolutely. The open problem is to obtain the constant Fourier mode of the seed function $\hat \Lambda _\Gamma$ for arbitrary loop order and weight. To any given order, this problem may be solved by partial fraction decomposition of the denominators, and use of the summation formula (\ref{Lif}).

\sm

We conclude by offering a speculation on the structure of the integrals of modular graph functions, defined above, for arbitrary loop order and weight.  It has been proven that the coefficients $\cL_k$ of the Laurent polynomial of $\cC_\Gamma$ are generated by multiple zeta-values with rational coefficients \cite{DHoker:2015sve}, and it was argued  that they are in fact generated by single-valued multiple zeta-values \cite{Zerbini,DHoker:2015sve}. We speculate that  the integrals are generated by multiple zeta-values (times a trivial factor of $\pi$ due to the volume of the fundamental domain), and  derivates with respect to the arguments of  zeta-values and multiple zeta-values.

\end{document}